# Understanding Data Better with Bayesian and Global Statistical Methods

William H. Press

Harvard-Smithsonian Center for Astrophysics

April 16, 1996

**Abstract**

To understand their data better, astronomers need to use statistical tools that are more advanced than traditional "freshman lab" statistics. As an illustration, the problem of combining apparently incompatible measurements of a quantity (the Hubble constant, e.g.) is presented from both the traditional, and a more sophisticated Bayesian, perspective. Explicit formulas are given for both treatments.

## 1 Introduction

Understanding data better is *always* an unsolved problem in astrophysics, although perhaps not in exactly the sense intended by the conference organizers. While other papers in this volume are more specifically directed at individual sub-areas of astrophysical theory, my contribution is intentionally more longitudinal: I hope that it is applicable to *all* the other areas surveyed.

If the spirit of this volume is to present a menu – a movable feast, indeed – of opportunities for thesis projects of smart second-year graduate students, then the opportunity that I would like to offer is one of voluntary self-choice: *Whatever* your choice of area, make the choice to live your professional life at a high level of statistical sophistication, and not at the level – basically freshman lab level – that is the unfortunate common currency of most astronomers. Thereby will we all move forward together.

What do I mean by "freshman lab level" and what do I mean by "sophisticated methods"? In my conference talk, I illustrated with three examples, in each case contrasting an elementary with a more sophisticated framework: chi-square fitting of parameters to a model; estimating correlation functions (in the simplest diagonal case, "error bars") from a data set; and combining independent, and perhaps incompatible, experimental measurements. Here, I will limit my discussion to the third topic only, both because I want to give a somewhat greater level of detail than was possible in the talk (enough detail to actually be useful in practice), and because my collaborative work on the other two topics is, or will be, written up elsewhere [14, 15, 12, 13].

## 2 Combining Experimental Measurements

We are given a number of supposedly independent measurements of a quantity, say, $H_1, \ldots, H_N$, with error bars on each one, $\sigma_1, \ldots, \sigma_N$. We are asked for the best estimate of the underlying quantity, call it $H_0$. (If this notation slyly reminds you of the Hubble constant, you are right!)

The conventional ("freshman lab") answer is to construct an average, weighted by the inverse variance of the individual observations,

$$H_0 = \frac{\sum_{i=1}^{N} H_i/\sigma_i^2}{\sum_{i=1}^{N} 1/\sigma_i^2} \qquad (1)$$

Equation (1) can be derived in any number of ways. For example, it is the maximum likelihood estimator in the case where each measurement has a normal distribution.

If you had a fairly advanced freshman lab, you also learned the formula for the variance of the estimator $H_0$, namely,

$$\sigma_0^2 = \frac{1}{\sum_{i=1}^{N} 1/\sigma_i^2} \qquad (2)$$

(If all the $\sigma_i$'s are equal, this says that the combined standard deviation is $1/\sqrt{N}$ times the individual standard deviations.)

You can derive equation (2) from equation (1) yourself simply by applying the Var( ) operator to equation (1) and using the rules

$$\text{Var}(\alpha x) = \alpha^2 \text{Var}(x) \qquad (3)$$

and
$$\text{Var}(x + y) = \text{Var}(x) + \text{Var}(y) \tag{4}$$
where $\alpha$ is a constant, and $x$ and $y$ are independent random variables. The variance that comes out is, of course, $\sigma_0^2$. The variance operator Var() is the usual one, defined by
$$\text{Var}(x) = \left\langle (x - \langle x \rangle)^2 \right\rangle \tag{5}$$

No matter how good your freshman lab instructor, it is almost a sure bet that he or she didn't show you the even more important formula for testing whether the individual measurements *are in fact compatible*, namely,

$$\chi^2 = \sum_{i=1}^{N} \frac{(H_i - H_0)^2}{\sigma_i^2} \tag{6}$$

If this value of $\chi^2$ is *not* compatible with $N - 1$ degrees of freedom (i.e., far *outside* the range $(N - 1) \pm \sqrt{2(N - 1)}$), then the estimate $H_0$ of equation (1) has simply no justification at all. The input values $H_i$ are simply incompatible.

Equation (6) derives from the fact that a chi square variable is the sum of squares of (zero-mean, Gaussian) quantities divided by their respective variances. However, you also have to know that you are supposed to treat $H_0$ as having zero variance, and instead reduce the number of degrees of freedom from $N$ to $N - 1$. (Somewhere around this point, rules of thumb give way to the better procedure of actually proving theorems. See [8], §§4.1 and 11.4.)

Currently, a generally recognized example of an incompatibility between multiple measurements and their respective claimed error bars is the Hubble constant. This has engendered a sometimes ferocious, and not always fact-based, debate about "which measurements to throw out". Of course it would be best to understand the physical basis for incompatibility between differing measurements. However, it is also true that, even absent such understanding, we do not have to throw all statistical analysis out the window. We can instead, as we will next illustrate, construct a well-posed statistical framework that allows apparently incompatible measurements to be combined in a useful way.

# 3 Bayesian Combination of Incompatible Measurements

We again have independent measurements $H_1, \ldots, H_N$, with claimed error bars on each one, $\sigma_1, \ldots, \sigma_N$. (We will call this, collectively, the data "$D$".) But now we want to be sophisticated enough to recognize that *some of the error bars are wrong*, due to (for example) unrecognized systematic effects, or unjustified optimism on the part of the observer. Is there any sensible, yet formal, way to sort all this out?

Here is a method that Chris Kochanek and I have worked out, with an unabashedly Bayesian derivation:

Suppose that $p_i$ is the probability that measurement $i$ is "correct" (in the sense of having accurate error bars), so that $1 - p_i$ is the probability that it is wrong (including the possibility that there are systematic errors, non-negligible in comparison to the quoted error bars). In this section we consider the model that all the $p_i$'s are the same, $p_i = p$. That is, $p$ is the "community-wide probability of doing a correct observation" at a certain epoch and in a certain field of science.

Let $\mathbf{v}$ be a vector of length $N$ whose $i$th component is either one or zero, signifying that the $i$th experiment is correct or wrong, respectively. Of course we don't know either the $p$ or $\mathbf{v}$ a priori.

There are three laws of probability that we will need to apply repeatedly. I can never remember their conventional names, so I will instead use descriptive unconventional ones. All three laws can be "proved" by drawing careful Venn diagrams, or by using the "frequentist" definitions for probability

$$P(A) \equiv \frac{\text{Number of Events with Property A}}{\text{Total Number of Events}} \qquad (7)$$

and for conditional probability

$$P(A|B) \equiv \frac{\text{Number of Events with Properties A and B}}{\text{Number of Events with Property B}} \qquad (8)$$

The laws are:

### 3.0.1 Law of Anding

$$P(ABC) = P(A)P(B|A)P(C|AB) \qquad (9)$$

That is, the probability of A and B and C is the product of three terms: the probability of A alone, the conditional probability of B *given A*, and the conditional probability of C *given A and B*.

### 3.0.2 Law of De-anding

If a set of hypotheses $B_i$ are exhaustive and disjoint, so that $\sum_i P(B_i) = 1$, then

$$P(A) = \sum_i P(AB_i) \qquad (10)$$

That is, we can recover the total probability of A from the sum of the more restrictive probabilities of *A and $B_i$*.

### 3.0.3 Bayes' Rule

If the set $\{H_i\}$ is an exhaustive and disjoint set of hypotheses, while $D$ is some data, then

$$P(H_i|D) \propto P(D|H_i)P(H_i) \qquad (11)$$

where $P(H_i)$ are the prior probabilities of the hypotheses (that is, their probabilities before the new data $D$ was gleaned). The constant implicit in the proportionality sign is determined by requiring the sum of the left-hand side (over $i$) to be unity. (One often saves effort in probability calculations by computing proportionalities and leaving the overall normalization to the end.)

In our problem, a complete hypothesis is a particular value $H_0$ *and* a particularly value for $p$ *and* a particular assignment of ones and zeros in $\mathbf{v}$. If, however, our primary interest is in $H_0$, then we use the law of de-anding and get

$$P(H_0|D) = \sum_{p,\mathbf{v}} P(H_0 p\mathbf{v}|D) \qquad (12)$$

The (formal) sum over $p$ of course becomes an integral if $p$ is (as it is) a continuous variable; the sum over $\mathbf{v}$ denotes a discrete sum over all $2^N$ possible values of the vector.

Next apply Bayes' Rule,

$$P(H_0|D) \propto \sum_{p,\mathbf{v}} P(D|H_0 p\mathbf{v}) P(H_0 p\mathbf{v}) \qquad (13)$$

and the law of anding,

$$P(H_0|D) \propto \sum_{p,\mathbf{v}} P(D|H_0 p\mathbf{v})P(H_0)P(p|H_0)P(\mathbf{v}|H_0 p) \qquad (14)$$

The four factors in (14) are now individually tractable. The first factor, $P(D|H_0 p\mathbf{v})$ is the probability of the data *given* (as $\mathbf{v}$) which experiments are right or wrong. For independent experiments one might model this as something like

$$P(D|H_0 p\mathbf{v}) = \prod_{v_i=1} P_{Gi} \prod_{v_i=0} P_{Bi}$$
$$= \exp\left[\sum_{v_i=1} \frac{-(H_i - H_0)^2}{2\sigma_i^2}\right] \times \exp\left[\sum_{v_i=0} \frac{-(H_i - H_0)^2}{2S^2}\right] \qquad (15)$$

Here $P_G$ and $P_B$ are the probability distributions of "good" and "bad" measurements, respectively. $S$ is a large (but finite) number characterizing the standard deviation of "wrong" measurements (e.g., plausible range in which a wrong measurement could have survived the refereeing process). Notice that since we are given $\mathbf{v}$, there is no additional information in $p$, so it does not appear on the right-hand sides of equation (15).

In other contexts, the use of Gaussians in an equation like equation (15) might prompt squeals of horror from the illuminati: Don't Gaussians always underestimate the tail probabilities of real measurements? And won't subsequent results therefore be quite sensitive to the Gaussian assumption? While the answer to the first question is of course "yes", the answer to the second is in fact "no". As long as $S$ is chosen to be adequately large, equation (14) will never be dominated by factors far out on the tails of the $\sigma_i$'s, because the dominating probability soon comes from the terms in the sum where the component of $\mathbf{v}$ is zero (so that $S$, rather than $\sigma_i$ is controlling). Relative insensitivity to tail probabilities is one of the appealing features of this formulation.

The second and third factors in equation (14) are our priors on $H_0$ and $p$. It is unlikely that our prior on $p$ (the probability of a typical experiment being "correct") depends on the value of, say, the Hubble constant, so, in fact, $P(p|H_0) = P(p)$. (The issue is not whether a particular experiment's chance of correctness depends on $H_0$ – obviously it does – but, rather, whether the single value $p$ that characterizes the current state of experimentation generally is somehow coupled to the expansion rate of the Universe – which seems unlikely!)

The fourth factor is also, on inspection, trivial: Given $p$, the probability of a particular value for the vector $\mathbf{v}$ is just the product of a factor of $p$ for each 1 component times a factor of $(1-p)$ for each 0 component.

Rearranging the summations, we can now rewrite equation (14) as

$$\begin{aligned} P(H_0|D) &\propto P(H_0) \sum_p P(p) \sum_{\mathbf{v}} [\prod_{v_i=1} P_{Gi} p][\prod_{v_i=0} P_{Bi}(1-p)] \\ &\propto P(H_0) \sum_p P(p) \prod_i [p P_{Gi} + (1-p) P_{Bi}] \end{aligned} \quad (16)$$

The last proportionality actually sums over all $2^N$ possible values of $\mathbf{v}$ and turns a $2^N$ computational problem into an $N$ one! (It is with some shame that I admit to having done some actual computer calculations, with $N = 15$, before this simplification was forcefully pointed out to me by Kochanek.)

Equation (16) is a complete, computationally feasible, prescription for calculating the probability distribution for the desired value $H_0$ given the mutually incompatible data. With no other information, one can take the priors $P(H_0)$ and $P(p)$ to be uniform, and compute just the indicated sum (actually an integral since $p$ is continuous) over the indicated products. (If you have greater a priori faith in your experimental colleagues, you can, of course, take a prior distribution for $P(p)$ that is more skewed towards higher values of $p$.) There is some judgment involved in the choice of $S$ in equation (15), but over a wide range the answers are typically insensitive to that choice.

One can see that the method is something like a maximum likelihood method that attributes to each measurement not a Gaussian, but a weighted sum (with weights $p$ and $1-p$) of a "good" Gaussian and a "bad" Gaussian. What makes the method manifestly Bayesian, however, is that it then integrates over all values of $p$, weighted by the prior $P(p)$. The traditional equation (1) is the mean of the distribution obtained in the limit that $P(p)$ is a delta function at $p = 1$, and with a uniform prior $P(H_0)$.

You can also go back and sum the four inner factors in equation (14) over $H_0$ and $\mathbf{v}$ to get the probability distribution for the variable $p$, or you can sum over $H_0$, $p$, and those $\mathbf{v}$'s with $v_j = 1$ (or $v_j = 0$) to get the probability that experiment $j$ is "good" (or "bad"). It is left as an exercise for the reader to derive the simpler forms analogous to equation (16) for these cases. In all cases, you get the constant in front by demanding that probabilities sum to 1.

Another easy exercise is to write down the formula for the probability that *none* of the experiments is good, that is, the total probability in the $\mathbf{v} = 0$ vector component. A value $> 0.05$ then indicates that *no* result is supported at the 5% confidence level minimally required for the reporting of scientific results.

Equation (16) is *not* a "tail-trimming" scheme that simply throws away outlier measurements. In some cases of actual data, the probability $P(H_0|D)$ will be multi-modal, with bumps of probability "protecting" certain outliers. In other cases, where there is a sufficient central core of mutually reinforcing values with mutually compatible error bars, the outliers will not be so protected. I have played around with the method on several data sets, including both published Hubble constant measurements (see Section 5, below) and measurements of $R_0$ (the distance to the Galactic center, [11]). In my experience the method is robust, and it gives results that are justifiable in terms of common sense. [1]

One important feature of the method is worth pointing out explicitly: Suppose there is a body of consistent results clustered around one value, but also a maverick outlier at an inconsistent value. In the conventionally weighted average of equation (1), the maverick is able to draw the average as close as he wants to his or her value, no matter how good the evidence on the other side, simply by publishing unrealistically small error bars. By contrast, with equation (16), smaller error bars (after a certain point) yield *decreasing* weight for the maverick value, because it becomes a "bad" data point with increasingly high probability. There are good statistical, and also good sociological, reasons for adopting a combining procedure with this characteristic.

It is of course true that, for the formulation of the problem given in this paper, a maverick can "stuff the ballot box" by repeated, mutually consistent (but wrong) measurements. Eventually those values would prevail. This is because we have not allowed explicitly for the possibility of correlated systematic errors in different experiments. A good problem for a graduate student would be to generalize the method (e.g., given some correlation matrix for the systematic errors of different experiments) so that repeated measurements, if highly correlated in their systematics, get only a single "vote". We have not done this.

---

[1] For $R_0$, I get $7.7 < R_0 < 8.4$ (Kpc) for the 95% confidence interval, and $7.9 < R_0 < 8.2$ as the 50% confidence interval. For $H_0$, see Section 5.

## 4 Another Variant of the Method

Kochanek (private communication) has pointed out a variant method: Instead of assuming a single probability $p$ applicable to the universe of experiments, one assumes an individual (though still unknown) probability $p_i$ for each experiment, that is, a vector $\mathbf{p}$. Then, instead of equation (16), we have

$$P(H_0|D) \propto P(H_0) \sum_{\mathbf{p}} \prod_i P(p_i|H_0)[p_i P_{Gi} + (1-p_i)P_{Bi}] \qquad (17)$$

Here the sum over $\mathbf{p}$ is actually a multidimensional integral over each of the $p_i$'s individually, but each applies to only a single term in the product, so

$$P(H_0|D) \propto P(H_0) \prod_i \left\{ \int_0^1 dp_i P(p_i)[p_i P_{Gi} + (1-p_i)P_{Bi}] \right\} \qquad (18)$$

If the priors on $p_i$ are a simple function, then the integrals can be done explicitly. Uniform priors, for example, give the exceptionally simple result

$$P(H_0|D) \propto P(H_0) \prod_i \frac{1}{2}(P_{Gi} + P_{Bi}) \qquad (19)$$

Comparing this with equation (16) one sees that the difference is that the integral over $p$ outside the product has been replaced by the average of $p$, namely 1/2, inside the product.

Kochanek and I have debated whether equation (16) is superior to equation (19) or vice versa. The conceptual difference is that, with all experiments sharing a single value of $p$, equation (16) is able to use their mutual compatibility or incompatibility to estimate $p$ in a non-trivial way, for example concentrating the probability distribution for $p$ strongly near 1 for highly concordant measurements. Equation (19), by contrast, has no information on any $p_i$ other than the prior, so it in effect uses $p_i \sim 1/2$. However, in numerical trials on actual data, there is not much difference in the inferred best-estimate values of $H_0$ for results of the two methods.

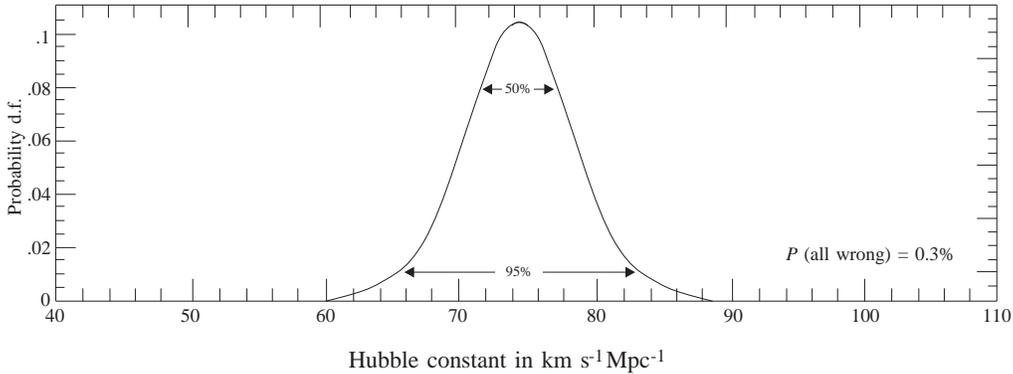

Figure 1: Bayesian probability distribution for the Hubble constant, derived from 13 different observations.

## 5   Results for the Hubble Constant

Although the intent of this paper is to discuss the statistical method, rather than the details of its application to the problem of the Hubble constant, it is probably a good idea for me to show some concrete results. Figure 1 shows the probability distribution for the Hubble constant $H_0$ that derives from equation (15) with $S = 30$ and a set of 13 reputable measurements of $H_0$ taken from the literature. The included measurements were chosen to include a variety of different techniques (including Type Ia supernovae, Type II supernovae, novae, globular clusters, Sunyayev-Zel'dovich effect, surface brightness fluctuations, planetary nebulae, Virgo cepheids, Tully-Fisher, and $D_N - \sigma$) and – where controversy exists – the values from more than one competing group. While there are surely correlations among the systematics in several of the measurements (notably the cepheid calibration), there are also methods that are entirely uncorrelated (e.g., Type II supernovae). See [3] for some further details.

One sees that a reasonably narrow distribution is obtained, with a 50% confidence interval of $72 < H_0 < 77$, and 95% confidence interval $66 < H_0 < 82$ (km s$^{-1}$ Mpc$^{-1}$). The probability that all of the observations are wrong (vector $\mathbf{v} = 0$) is 0.3%, a comfortingly small value. The superficial resemblance of the result to a Gaussian is a result, not an input constraint.

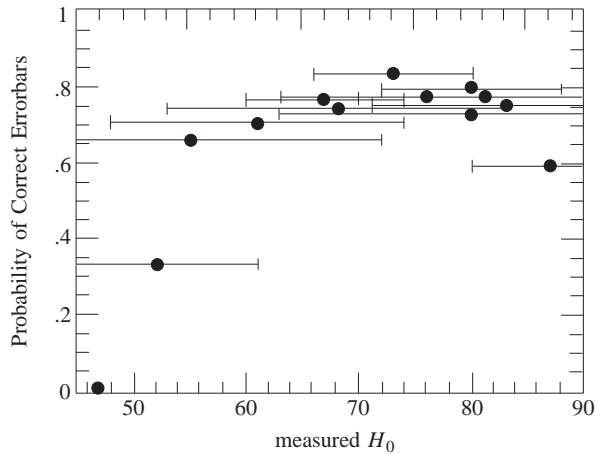

Figure 2: Projected onto the horizontal axis, this figure shows the values and error bars that were used to obtain the results of Figure 1. The vertical axis gives, for each measurement, the posterior estimate that it is "correct" (that is, compatible with its own error bars given the weight of evidence of all the other measurements).

(Indeed, if one arbitrarily decreases the quoted error bars of the experiments by a factor of 3, the "Gaussian" breaks up into a tri-modal distribution with "low", "medium", and "high" values for $H_0$.)

The abscissa of Figure 2 shows the input values and error bars assumed for the 13 measurements. (I am intentionally omitting a detailed list of references to avoid, or at least evade, an outburst of controversy. The point is to illustrate a statistical technique, not give a definitive review of $H_0$.) The ordinates of Figure 2 are the posterior estimates for the probability that each measurement has its $v_i = 1$ (is "right") rather than $v_i = 0$ (is "wrong"). One sees that most measurements are estimated to have about a 75% chance of being "right".

Finally, Figure 3 shows the posterior estimate for $P(p)$, the probability distribution for the prior probability of a random $H_0$ measurement being correct. It peaks at around 75%, but has a significant tail extending to zero. Note that the Bayesian nature of our method makes it, in effect, an *average*

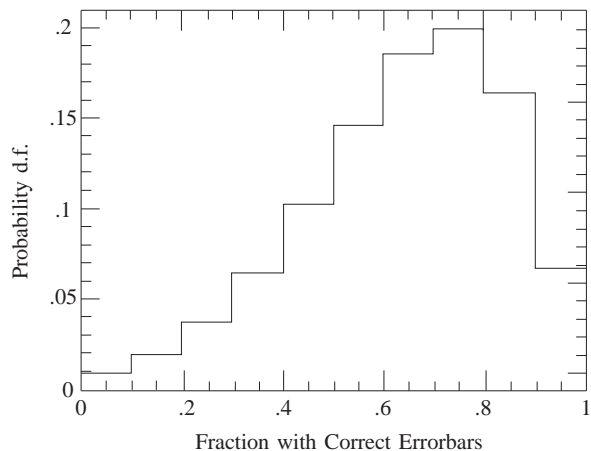

Figure 3: Figure 3. Posterior distribution for $p$, the prior probability that a given measurement of the Hubble constant is correct.

over this distribution of $p$, rather than assume any single value.

# 6    Conclusion

John Bahcall, a friend and mentor to me for more than 20 years, has always spurred those around him to choose important problems (which he often himself suggests), to get involved with real data (which he often himself supplies), and to bring the most powerful analytic tools to bear on the problem at hand. This paper is addressed at the last of these three imperatives. John's own work sets a mighty standard in all three respects, and his work will be studied for effective techniques and important conceptualizations long after the specific data are (as they should be) obsoleted by future observational advances.

# Bibliographic Notes

Three good starting points for astrophysicists who want to learn some statistics, at a not totally unsophisticated level, are [8], [2], and [9], the latter two (alas) now somewhat out of date.

A good introduction to Bayesian inference, from an astronomical perspective is [7]. Several other papers in the same volume [4] will also be of interest. If you are a doubter who thinks that there is a "subjectivity" in Bayesian statistics that is not present in conventional frequentist approaches, you should look at [1], which shows that the equivalent (or worse) subjectivity is present in conventional approaches and is merely better camouflaged.

There is a series of conferences on "Maximum Entropy and Bayesian Methods" whose proceedings frequently contain papers of interest to astronomers. Recent volumes are [16] and [10].

An astronomical area in which Bayesian analysis has recently shown itself to be powerful is the determination of the mass of the Milky Way from orbital information on satellites. See [6] and, more recently (with some of the same viewpoints as this paper), [5].